\documentclass[12pt]{article}
\usepackage{amssymb,amsmath}
\usepackage{array}
\usepackage{float}
\usepackage{graphicx}
\usepackage{xcolor}

\def\'#1{{\accent19\ifx #1i \i\else #1\fi}}

\def\v#1{{\boldmath#1}}

\def\be{\begin{equation}}
\def\ee{\end{equation}}
\def\bea{\begin{eqnarray}}
\def\eea{\end{eqnarray}}

\newcommand{\boldmathPhi}{\mbox{\boldmath$\Phi$\unboldmath}}

\newcommand{\boldmathsigma}{\mbox{\boldmath$\sigma$\unboldmath}}
\newcommand{\boldmaththeta}{\mbox{\boldmath$\theta$\unboldmath}}

\newcommand{\boldmathtau}{\mbox{\boldmath$\tau$\unboldmath}}

\newcommand{\boldmathPsi}{\mbox{\boldmath$\Psi$\unboldmath}}
\newcommand{\boldmathI}{\mbox{\boldmath${\it I}$\unboldmath}}

\newbox\Ancha
\catcode`@=11
\newdimen\ex@
\title{Heavy-quark   mass relation from a standard-model boson   operator  representation in terms of fermions}
\author{Jaime Besprosvany and Rebeca S\'anchez }
\date{Instituto de F\'{\i}sica, Universidad Nacional Aut\'onoma de M\'exico,
Apartado Postal 20-364,  01000, Ciudad de M\'exico, M\'exico }

\begin{document}

\maketitle









\jot = 1.5ex
\def\baselinestretch{1.10}
\parskip 5pt plus 1pt


\begin{abstract}

The standard-model can   be equivalently  represented  with its  fields       in a spin-extended basis,   departing from fermion degrees of freedom. The common Higgs operator  connects the electroweak and Yukawa sectors, restricting the top and bottom quark masses[Phys. Rev.  D {\bf 99}, 073001, 2019].
Using   second quantization,   within the   heavy-particle  sector,   electroweak vectors,  the Higgs field,  and  symmetry operators  are expanded
  in terms of bilinear combinations of  top and  bottom quark    operators, considering  discrete degrees of freedom and chirality.     This is interpreted as either a basis choice or as a description of  composite models.
    The   vacuum expectation value
 is calculated  quantum mechanically, which relates to
the  common   mass-generating scalar operator and it reproduces  the  vector  and    quark-doublet masses.
  This also links the  corresponding  scalar-vector and Yukawa vertices,
  and
restricts    the t-  and  b-quark masses  in a hierarchy relation.
\end{abstract}
\baselineskip 22pt\vfil\eject \noindent

\renewcommand{\thesection}{\Roman{section}}

\section{Introduction}




The standard model\cite{Glashow} (SM) is pivotal in obtaining information about elementary particles, yet     theoretical puzzles remain as the  connection between the electroweak,  Higgs and Yukawa sectors, and the origin of the electroweak breaking. New connections may lead to understanding of     SM dimension and dimensionless  variables that are  fixed phenomenologically today.

The SM's properties provide hints to address these puzzles. Quantum numbers of SM spin-1/2 and vector particles are related, as these belong to the fundamental and the adjoint representations, respectively,
of  the gauge and Lorentz symmetry groups; to implement such symmetries,  the vector bosons and the Higgs scalar have charge and spin of specific pairs of SM fermions.

One productive approach to obtain new information on the SM relies on these patterns. Ref. \cite{BesproRom} rewrites  the SM Lagrangian equivalently in terms of a matrix basis,  for all particles (scalar, vector, spin 1/2). It    derives a mass relation (classically) for the quark masses, under a weak SM assumption of a common scalar operator between the electroweak sectors,  using a discrete equivalent basis induced by a SM extension; spin and gauge degrees of freedom are separated, and the bosons’ degrees of freedom are formally composite of the fermions.
This suggests the  possibility of a SM spin-1/2 basis and associated quantum numbers, under the weak assumption of truncation that can be tested.

On the other hand, SM extensions with a bottom-up approach,  generalizing  SM aspects,   have   led to restrictions on these parameters, beyond the insight  of an encompassing theory. For example, grand-unified theories\cite{Georgi}  assume a common group for the interactions, requiring a unique coupling constant at the unification scale,  constraining    the SM  couplings.

 The discrete composite SM quantum numbers suggest   such a  description of the elementary particles, and hint at   generalization directions. Compositeness is a central tenet  behind many physical systems, in which features are  explained in terms of simpler  elements.  It means   a  system's configurations  are divided into  two classes,      elementary   and composite,   where the latter's  degrees of freedom  are constructed in terms of the former's, and   observables  manifest these relations.  In the case of the quark model\cite{quarkmodel},     hadron masses are explained in terms of constituent-quark masses. In superconductivity, Cooper pairs\cite{Cooper}, produced by a residual phonon interaction   between  two electrons, conform the relevant free-streaming  variable.

Thus, in  the BCS theory of superconductivity\cite{Schrieffer}    the Cooper-pair energy gap  depends on  interactions. In an application  of this theory to  quantum field theory and elementary particles\cite{NambuJona},
 a  four-fermion   interaction   produces fermion     and composite-boson   masses,
 linking  their values.

 Based on this type of interaction, applying dynamic mass generation, models were researched\cite{Miransky} that produce  a Higgs-scalar condensate\cite{topcondensate},  composed of top and antitop quarks. These models comprise, among others, an extension to vector particles\cite{MiranskyMassWZ}, applying interaction resummation\cite{Ross},
  with consideration of a chiral condensate\cite{chiralconden}.


In finding the  appropriate description of a composite system, the focus is on  effective degrees of freedom, concentrating the relevant information.
In quantum field theory, fundamental variables as   coupling constants and masses run, depending on the energy scale, as renormalization  equations   take  into account  interactions.

 Once these properties are integrated, a composite configuration  has a simple description in terms of the elementary ones.
Compositeness may be present at a fundamental level or partially,  depending on whether it involves  all or selected number of the   relevant degrees of freedom.
 SM  compositeness  features  may prove useful to address the   SM problem    that the    fermion sector  remains  disconnected from the boson  elements, and so the masses that they generate, which arise  from independent Lagrangian terms.


In this paper,  we  derive a  mass relation for the heavy quarks  in terms of electroweak parameters, and a hierarchy constraint,  relying  on SM compositeness, using the common mass-generating Higgs component, in a quantization framework.     Concentrating on SM heavy particles,     boson degrees of freedom can be expressed   in terms of the  fermions',  considering spin and  isospin.     From  the electroweak   and  Yukawa    Lagrangian components,   the vector Z and W masses   are reproduced. The material is organized thus: in Section II,   mass-generating operators are extracted from the electroweak Lagrangian. Section III relates the paper scheme to composite top quark anti quark Higgs condensate models. In Section  IV, such operators are written in a second-quantized  fermion basis; these are used to calculate particle masses and derive the quark hierarchy constraint, in Section V.  Conclusions are drawn in Section VI.

\section{Vector-scalar  and fermion-scalar mass components }

The SM assumes a  Higgs doublet
\begin{eqnarray} \label{HiggsReBase}{\bf H}(x)=\exp[{i\boldmaththeta(x)\cdot\boldmathtau}/(2 v)] \left [\frac{1}{\sqrt{2}}\left ( \begin{array}{c}  0 \\
  h(x)\end{array}  \right )+ {\bf v}\right ], \end{eqnarray}  parameterized in terms of $\boldmaththeta(x$) phase   fields, with Pauli-matrix   generators ${\boldmathtau}$,    and the vacuum expectation value (vev)
${\bf v}= \frac{v}{\sqrt{2}}\left ( \begin{array}{r}  0 \\
  1 \end{array}  \right )$; through the Higgs mechanism, an expansion around the scalar potential minimum produces
$h(x)$ as the remaining only physical  field, which minimizes the potential.  In the unitary gauge, the $\boldmaththeta(x$)   are absorbed into the vectors' longitudinal components, and  the exponential  dependence  is eliminated.


 Next, we    write the scalar-vector  (SV) Lagrangian
 \begin{eqnarray} \label{SVmassMEPrime}
 {\cal L  }_{SV}={\bf H}^\dagger(x) \left  [ \frac{1}{2}g  {\boldmathtau}\cdot {\bf W^\mu}(x)+\frac{1}{2}g'B^\mu(x)\right ]  \left  [ \frac{1}{2}g  {\boldmathtau}\cdot {\bf W_\mu}(x)+\frac{1}{2}g'B_\mu(x)\right ]   {\bf H}(x) , \end{eqnarray} where  ${W^i}_\mu(x)$, $i=1,2,3,$ are   weak vector components,  ${B}_\mu(x)$    is the hypercharge  vector term, and  $g$, $g '$ are the respective coupling constants. In the Higgs mechanism,  ${\bf H}(x)$ acquires a    vev    $\langle{\bf H}(x)\rangle={\bf v}$, producing  the  SM mass-generating  constant component,  contained in   ${\cal L  }_{SV}$:    \begin{eqnarray} \label{SVmassME}
    \nonumber
 { \cal L}_{MV}= \frac{1}{4} {\bf v}^\dagger \left  [g  {\boldmathtau}\cdot {\bf W^\mu}(x)+g'B^\mu(x)\right ]\left  [     g  {\boldmathtau}\cdot {\bf W_\mu}(x)+ g'B_\mu(x)\right ] {\bf v} , \end{eqnarray} which  provides masses to the vectors.
  The fields' quadratic form is resolved in
  \begin{eqnarray} \nonumber
 { \cal L}_{MV}= \frac{1}{{g }^2 +{g'}^2} {\bf v}^\dagger \left  [  ( {g }^2  I^3-\frac{1}{2}{g'}^2 Y)Z^\mu(x) \right ]\left  [    ( {g }^2  I^3-\frac{1}{2}{g'}^2 Y)Z_\mu(x)\right ]
    {\bf v}+& \\
  {\bf v}^\dagger\left  [\frac{1}{\sqrt{2}}g     I^-  { {W^-}^\mu}(x)\right ]\left  [   \frac{1}{\sqrt{2}}g   I^+  { {W^+}_\mu}(x)   \right ]  {\bf v}= \\  \label{SVmassMEZA}
  \frac{1}{2} Z_\mu(x)Z^\mu(x)   M^2_{Z}+   {W^-}_\mu(x){W^+}^\mu(x)  M^2_{W},
 \end{eqnarray}
where the weak isospin generators are $\frac{1}{ 2 }{\boldmathtau}={\boldmathI}$, with $I^\pm= I^1\pm i  I^2 $,   and  $Y$ is the hypercharge operator. This form is expressed in terms of the charged
  $ {W^\pm}_\mu (x)=\frac{1}{\sqrt{2}}[ W^1_\mu(x)\mp i W^2_\mu(x)]$, and
neutral components that decouple in the  above quadratic form:
\begin{eqnarray}
 Z_\mu(x)&=&\frac{1}{\sqrt{{g }^2 +{g'}^2}}[-g W^3_\mu(x)+g' B_\mu(x)] \\
 A_\mu(x)&=&\frac{1}{\sqrt{{g }^2 +{g'}^2}}[g'  W^3_\mu(x)+g B_\mu(x)],
 \end{eqnarray}
where the photon field  $A_\mu(x)$ remains massless:
${g }g '(  I^3+ \frac{1}{2} Y) A_\mu(x) {\bf v}=0,$ whereas the obtained masses for the ${\rm Z}$ and ${\rm W}^\pm$ particles are
\begin{eqnarray}\label{particleMasses}
 M^2_{Z}&=& \frac{2}{{g }^2 +{g'}^2}{\bf v}^\dagger ( {g }^2  I^3-\frac{1}{2}{g'}^2 Y) ( {g }^2  I^3-\frac{1}{2}{g'}^2 Y){\bf v} = (g^2+{g'}^ 2)v^2/4  \\ \label{particleMassesN}
 M^2_{W}&=&\frac{1}{2}{\bf v}^\dagger I^-   I^+ {\bf v}= g^2 v^2/4 .
 \end{eqnarray}


We focus on the heavy-quark  chiral-(t,b)   field components with left-handed doublet    \begin{eqnarray} \label{FermionDoublet}{ \boldmathPsi}_L(x)= P_L  \left ( \begin{array}{c}  \psi_t(x) \\
 \psi_b(x)\end{array}  \right ),\end{eqnarray} and right-handed singlets $\psi_{tR}(x)  =P_R\psi_t(x)$, $\psi_{bR}(x)  =P_R\psi_b(x)$,   with the projectors $P_R =\frac{1}{2}(1+  \gamma_5), $ $P_L =\frac{1}{2}(1-  \gamma_5). $ The scalar-fermion (Yukawa) interaction component is
\begin{eqnarray}\label{LagFerMass}{\cal L}_{SF}= \chi_t{\bar\boldmathPsi_L}^\dagger (x) \tilde {\bf H}(x) \psi_{tR}(x)+ \chi_b  {\bar\boldmathPsi_L}^\dagger (x)   {\bf H}(x)  \psi_{bR}  (x)    +{\rm H. c.},
\end{eqnarray}
where $\chi_t$, $\chi_b$ are Yukawa constants, and
\begin{eqnarray}\label{Htilde}\tilde {\bf H}(x) = i\tau^2 {\bf  H}^*(x). \end{eqnarray}

The extracted mass-generating term from ${\cal L}_{SF}$, after the Higgs mechanism, is

\begin{eqnarray}\label{LagFerMassAfterHiggsMec} {\cal L}_{fM}= \chi_t \frac{v}{\sqrt{2}} \bar  \psi_{tL}(x)  \psi_{tR}(x) +\chi_b \frac{v}{\sqrt{2}}\bar\psi_{bL}(x)  \psi_{bR}(x) +{\rm H. c.}
\end{eqnarray}
  As suggested by the Lagrangian components in  Eqs.  \ref{particleMasses}, \ref{particleMassesN}, \ref{LagFerMassAfterHiggsMec},  the fields   may be written   in terms of a fermion basis,
also indicated by the  fermion-vector interaction term.  Recalling the SM quantum-number
 compositeness property, from gauge invariance,   bosons are written in terms of fermions, as some vector-field zero components constitute  conserved terms.
We concentrate on relevant discrete degrees of freedom: spin and  isospin, in their chiral components, contributing to the masses, as implied by Eqs. \ref{particleMasses}, \ref{particleMassesN}, \ref{LagFerMassAfterHiggsMec}.   The  vector field   has the form\cite{Wess} $A_\mu= g_{\mu\nu}A^\nu= \frac{1}{4} {\rm tr} \gamma_\mu \gamma_\nu A^\nu $  in a spinor basis.  The SM Lagrangian  is   reformulated   in  a  spin-isospin   basis, in a similar procedure   to  the  spin-extended model\cite{BesproRom}.  The vector-scalar component leading to the vectors' masses is, from Eq. \ref{SVmassME},
 \begin{eqnarray} \nonumber
   {\cal L}_{SV}= {\bf H}^\dagger(x)[\frac{1}{2}g  {\boldmathtau}\cdot {\bf W^\mu}(x)+\frac{1}{2}g'B^\mu(x)][  \frac{1}{2}g  {\boldmathtau}\cdot {\bf W_\mu}(x)+\frac{1}{2}g'B_\mu(x)] {\bf H}(x) = \\ \nonumber
 \frac{1}{2} {\rm tr}{\bf H}^\dagger(x)  \gamma_0  [\frac{1}{2}g  {\boldmathtau}\cdot {\bf W ^\nu}(x)+\frac{1}{2}g'B^\nu(x)]   {\gamma_\nu}     [\frac{1}{2}g  {\boldmathtau}\cdot {\bf W^\mu}(x)+\frac{1}{2}g'B^\mu(x)]    \gamma_\mu   \gamma_0 {\bf H} (x) ,
 \end{eqnarray} \begin{eqnarray}  \label{SVmass} \end{eqnarray}
 where the trace here is only over $4\times 4$ gamma-matrix indices, and the equality uses
  \begin{eqnarray}\label{SVmassClifford} g_{\mu\nu}&=&\frac{1}{4} {\rm tr} \gamma_\mu \gamma_\nu = \frac{1}{4} {\rm tr} \gamma_0   P_L    \gamma_\mu \gamma_\nu P_L \gamma_0 + \frac{1}{4} {\rm tr}\gamma_0   P_R \gamma_\mu  \gamma_\nu P_R\gamma_0 ,  \\
   \label{SVmassCliffordDetail} &=&\frac{1}{2} {\rm tr}   \gamma_0  P_L \gamma_\mu  \gamma_\nu P_L  \gamma_0.   \end{eqnarray}


  The inclusion of the $P_R$, $P_L$ projections shows the freedom in the representation choice,
where  the independence of symmetry-operator spaces (spin, isospin) is manifest, and the tensor product of spaces applies for its generators. Eqs. \ref{SVmassClifford}, \ref{SVmassCliffordDetail} signal   a new basis that separates Lorentz and gauge degrees of freedom, satisfying the  Coleman-Mandula theorem\cite{Mandula}, similarly but independently of a proposed SM extension,   in Ref. \cite{BesproRom}.
 With hindsight,  the scalar doublet   ${\bf H}(x)$,   is written in the spin basis (using $\gamma_0$)
 while a full quantum version    in second quantization is
  provided in Section IV.

 Such a  formulation  introduces an equivalent basis with creation and annihilation operators  incorporating    spin and scalar (weak isospin, flavor) components as   relevant degrees of freedom, and their  vev.
   Next, we consider SM  limits and composite extensions for which the expansion in this paper is relevant.


\section{Standard-model limits and its composite extensions}

To face SM puzzles, as the origin of the electroweak breaking or the connection of the electroweak and Yukawa sectors, relevant limiting-environments are described for this work; in turn, its calculation   leading to the SM vector masses complements these methods.

\subsection*{Heavy-quark sector behavior}
The large-mass limit of SM heavy particles     was shown\cite{HeavyQuarks}   to conserve the SU(2)  symmetry. Using path integrals, quantum corrections to classical solutions are obtained, as the Higgs electroweak and Yukawa Lagrangians maintain their forms. Unlike Ref.  \cite{HeavyQuarks}, initial conditions may be set for the quark components with arbitrary chiral quarks $u_L$, $u_R$, $b_L$, $b_R$.
\subsection*{Chiral perturbation theory }
This theory\cite{ChiralPertubationT}  matches the paper’s fermion chiral operator relevant degrees of freedom; the choice in Eq. \ref{massquark} manifests the chiral symmetry is broken, as quarks are   particles with well-defined parity.

\subsection*{Effective theory}
The old Fermi theory gives an elementary description of the SM  particles and interactions in the low-energy limit, through   an effective theory,   in terms of fermions.  This is a parameter-reducing approach, connecting, e. g.,  bilinear scalar and vector components. So,   interactions may be added    using  only fermions, with various applications.

\subsubsection*{Nambu--Jona-Lasinio models}  These    interactions are constructed from   four fermions\cite{{NambuJona}}, which come  from various origins,  as in the low-energy SM description, in which they manifest a vector-boson exchange.

In the descripton of  SM heavy-particles, such an interaction term, satisfying the SM symmetries,  is chosen in terms of the heaviest fermions' fields,  top and bottom quarks:
 \begin{eqnarray} \label{NambuJonaSM}
{ \cal L}^{\Lambda}_{\rm old}={ \cal L}^{0}_{kin}+ G(
{\bar\boldmathPsi}_L^{ia} t_R^a )( \bar t_R^b {\boldmathPsi}_L^{
ib}),
  \end{eqnarray}
where $G$ is a coupling, $i$ is an electroweak index, $a$, $b$ are color indexes, ${ \cal L}^{0}_{kin}= \bar \boldmathPsi   \slash \!   \!  \!  \partial   \boldmathPsi$, and
 we use the $\boldmathPsi$ doublet in Eq. \ref{FermionDoublet} with an explicit color index, adapting the t,b notation.

\subsubsection*{Top-quark condensate models}
The heavy-particle SM sector suggests   common dynamics as an explanation to electroweak symmetry breaking, as this property implies these  particles are linked.  SM extensions comprise various schemes and methods that  maintain the SM electroweak structure, providing connections among SM variables, and we list some involving compositeness. In one such extension,    this entitles the introduction of a four-fermion attractive interaction  of Eq. \ref{NambuJonaSM},  generating,  through a Nambu--Jona-Lasinio mechanism\cite{NambuJona},   massive quarks, and a composite Higgs particle, whose main component is a top-antitop pair,    producing a    condensate in  various models\cite{ReviewTopCondensate}.

The original scalar doublet is represented by
 \begin{eqnarray} \label{HiggsNJ}
\tilde \boldmathPhi\ \alpha\ \frac{1}{\sqrt{2}}\left ( \begin{array}{c} \bar b ^a (1+\gamma_5)  t ^a \\
 -\bar t^a(1+\gamma_5) t ^a\end{array}  \right ) =i\tau_2 \tilde \boldmathPhi^*
  \end{eqnarray}


   \begin{eqnarray} \label{NambuJonaSMPhi}
 \tilde \boldmathPhi\ \alpha\ i\tau_2    \boldmathPhi^* = \bar t_R^b {\boldmathPsi}_L^b
  \end{eqnarray}

  An alternative formulation with the same physical contents is given by considering from Eq. \ref{NambuJonaSM}:
 \begin{eqnarray} \label{NambuJonaSMLinear}
{ \cal L}^{\Lambda}_{\rm new}&=&{ \cal L}^{\Lambda}_{\rm old}   -(
  M_0  {\tilde{ \boldmathPhi ^i }}^\dagger    + \sqrt{G}\bar\boldmathPsi_L^{ia} t_R^a )( M_0 \tilde {\boldmathPhi}^i+\sqrt{G} \bar t_R^b {\boldmathPsi}_L^{
ib})\\
&=& \bar \boldmathPsi   \slash \!   \!  \!  \partial   \boldmathPsi -  \sqrt{G} M_0 ( {\tilde{ \boldmathPhi ^i }}^\dagger  \bar t_R^b {\boldmathPsi}_L^{
ib} +\bar\boldmathPsi_L^{ia} t_R^a  \tilde {\boldmathPhi}^i) - M_0^2 {\tilde{ \boldmathPhi ^i }}^\dagger  {\tilde\boldmathPhi}^i,
  \end{eqnarray}
where  $M_0$  is a  mass parameter, and  we introduce the auxiliary fields
   \begin{eqnarray} \label{HiggsAuxiliary}
\tilde \boldmathPhi= \frac{1}{\sqrt{2}}\left ( \begin{array}{c} { \cal G}^{(1)}+ i{\cal G}^{(3)} \\
  { \cal H}+ i{\cal G}^{(0)}\end{array}  \right ) =i\tau_2   \boldmathPhi^*.
  \end{eqnarray}
 Three   conform Goldstone bosons that eventually are absorbed by  vector bosons.  leaving the physical Higgs.
The equations of motion from Eq. \ref{NambuJonaSMLinear} imply  Eqs. \ref{HiggsNJ}, \ref{NambuJonaSMPhi}, defining their constants.

This composite structure reproduces  the SM as a minimum-energy solutions. The   Higgs mechanism
implies a scalar composite scalar encompassing     a quark condensate of a composite Higgs,  manifesting  vev of the scalar real field \cite{Miransky} $\langle {\cal H}(x)\rangle =v$

A quark condensate\cite{topcondensate} is based on previous work\cite{Miransky,MiranskyMassWZ}
  that connected the NJL theory  to the renormalization group, and improved its predictions.
In  theory, within an  energy scale $\Lambda\sim$ 10$^8$ MeVs,
the renormalization group reveals that top quark condensation is fundamentally based upon the ‘infrared fixed point’ for the top quark Higgs-Yukawa coupling\cite{Pendleton,Hill}. The ‘infrared’ fixed point originally predicted that the top quark would be heavy, contrary to the prevailing view of the early 1980s.  Such a  point implies that it is strongly coupled to the Higgs boson at very high energies, corresponding to the Landau pole of the Higgs-Yukawa coupling. At this high scale a bound-state Higgs is formed  in the ‘infrared’, as the coupling thus relaxes to its measured value of order unity. The SM renormalization group fixed point prediction is about 220 GeV, as  the observed top mass is roughly 20\% lower than this prediction.
The simplest top condensation models are now ruled out by the LHC discovery of the Higgs boson at a mass scale of 125 GeV. However, extended versions of the theory, introducing more particles, can be consistent with the observed top quark and Higgs boson masses.

Either results are too restrictive, not in agreement with experiment or open to new processes, and parameters, losing predictability. Still, they provide a framework in which the below results can be formulated. Other approaches include the
direct Bether-Salpeter equation, introducing QCD corrections, technicolor, and extended Higgs doublets.

 The Higgs mechanism\cite{Higgs}  implies  massive-field configurations for vectors and fermions   as  the scalar field is expanded around the minimum energy state.
Massive  quarks  are Dirac particles, but their original chiral components can be used as a basis.
  Such   massless components maintain their quantum numbers but acquire mass.
In a mode expansion over momentum degrees of freedom, we need only look at the mass components.

The pervasive    original SM electroweak Yukawa masses framed in these  composite extensions, described the Higgs conditions  from the Higgs field, are a framework for the which the free-particle quantized description below is relevant.

\subsection*{Extensions and basis in a Lagrangian formulation }

 Beyond the SM (BSM) theories   can be tested, viewed as approximations to the SM,  through a   sytematic perturbative expansion,
involving the SM Lagrangian:
\begin{equation}\label{LagrangianU1}  {\cal L}_{SM}= {\cal L}_{BSM}+{\cal L}_{SM}-{\cal L}_{BSM}\end{equation} contains the  SM Lagrangian and ${\cal L}_{BSM}$       with through successive corrections, ${\cal L}_{SM}-{\cal L}_{BSM}$.

A second interpretation of the expansion is of a  different new basis that can be tested,   as it is obtained from the old one by
  \begin{equation} \label{basisExpansion}
    |N\rangle=
  e^  {i (\cal L _N-\cal L _O)}   |O\rangle, \end{equation}   providing a solution. We conjecture a Lagrangian
  that assumes an additive operator organization, thus, a unitary transformation  perturbation expansion can be constructed through   Hermitian operators
  $ \cal L _N+\cal L _O-\cal L _N$, suggesting an expansion in
 $ \cal L _O-\cal L _N$ corrections.  To  the extent the original SM is maintained,    gauge-invariance theorems, under the Higgs mechanism, based on a lattice description\cite{Frohlich},   hold.


\section{Second-quantized field expansion }
We use the  8-element massive  basis for the t quark   (and  the  b),
conformed of a tensor product of the spaces isospin, momentum,   spin, $q_{ks}= a_{q} a_{k} a_{s} $.   We concentrate on the top  and bottom quarks   $q=t,b$,   the up and down spin polarization, $i=\uparrow,\downarrow$, as these constitute the most massive
 fermions, with masses of the order of the  SM massive bosons. As we deal with a QCD-scalar SM sector, any quark color is understood. The momentum (or space) degrees of freedom are factored out; for definiteness,  among  the new components, we choose    the lowest-energy fermion operator, the massive mode, with  ${\bf k}\rightarrow 0,  $   as any    mode  is  representative, given the  Lorentz and gauge invariance.
Such objects satisfy the    anticommutation relations \begin{eqnarray}\label{anticommutationDirac}  \{   q_{i} ^\dagger, q'_{j} \}=\delta_{ij} \delta_{qq'}, \  \{ q_{i},q'_{j} \}=\{ q_{i}^\dagger,{q'}_{j}^\dagger\}=0,\end {eqnarray}
with Kronecker deltas  understood for defined  indices.
Quarks $ q_{i}^\dagger$ have the  same quantum numbers as antiquark operators $\bar q_{i}$, with  antiquarks   given by $\bar q_{i}   ^\dagger  $.    Action on the vacuum is    $q_{i}   |0  \rangle=0$, $\bar q_{i}   |0  \rangle=0$.   The  normalization is set for the massive states,
 $\langle  0|  q_{i}  q_{i}^\dagger|0  \rangle=\langle  0|  \bar q_{i}  \bar q_{i}^\dagger|0  \rangle=1 $.


 Given the chirality's admixture of positive and negative massive frequencies, these operators can be broken into their right and left components and are  represented\footnote{Strictly speaking, the massive ${\bf k\rightarrow 0}$ quark component is obtained from the  off-shell  massless    operators;  their ${ \bf k}$, ${-\bf k}$ redundancy  is eliminated in such a limit. For example, $t_{R\uparrow}^\dagger({\bf k})$ corresponds to $\bar t_{L\downarrow}({-\bf k})$.} by
 \begin{eqnarray}\label{massquark}q_{\uparrow}^\dagger= \frac{1}{\sqrt{2}}( q_{R\uparrow}^\dagger + q_{L\uparrow}^\dagger),\end {eqnarray}  and  the antiquark  annihilation operator, the orthogonal combination
  \begin{eqnarray}\label{massquarkdestru}\bar q_{\downarrow} = \frac{1}{\sqrt{2}}(  q_{R\uparrow}^\dagger - q_{L\uparrow }^\dagger),\end {eqnarray}
 and the hermitian and the spin $\uparrow\leftrightarrow\downarrow$ equations.  These relations imply for the quiral $L,R$ quark-basis components, using Eq. \ref{anticommutationDirac},
    the anticommutation relations  $\{ q_{Qi}^\dagger, q' _{Q'j} \}= \delta_{ij} \delta_{QQ'} \delta_{qq'}$,   and $Q,Q'=R,L$,    and for anticommutator of  creation-creation
     and annihilation-annihilation operators, zero.

\begin{table}[H]
\noindent \begin{raggedright}
\begin{tabular}{|>{\raggedright}p{0.21\textwidth}|>{\centering}p{0.07\textwidth} >{\centering}p{0.05\textwidth}>{\centering}p{0.05\textwidth} >{\centering}p{0.05\textwidth}  >{\centering}p{0.05\textwidth}                            >{\centering}p{0.05\textwidth} >{\centering}p{0.05\textwidth}  >{\centering}p{0.05\textwidth} |}
\hline
  operator &  $Q$ & $\hat k$  & $B$ & $Y$ &  $I^3$ & $\bar I^2$ &  $S_z$   &  $\bar S^2$  \tabularnewline
\hline
$t_{R\uparrow}^\dagger $  & $\begin{array}{r}
2/3
\end{array}$ & $\begin{array}{r}
1
\end{array}$  &
$1/3$
  & 4/3 & $\begin{array}{r}
0
\end{array}$ &$\begin{array}{r}
0
\end{array}$ & $\begin{array}{r}
1/2
\end{array}$
& $\begin{array}{r}
1/2
\end{array}$
\tabularnewline
$t_{R\downarrow}^\dagger $  &  $\begin{array}{r}
2/3
\end{array}$ & $\begin{array}{r}
$-1$
\end{array}$  &
$1/3$
  & 4/3 & $\begin{array}{r}
$0$
\end{array}$ &$\begin{array}{r}
0
\end{array}$ & $\begin{array}{r}
$-1/2$\end{array}$ & $\begin{array}{r}
$1/2$
\end{array}$\tabularnewline
$t_{L\uparrow}^\dagger $ &  $\begin{array}{r}
$2/3$
\end{array}$ & $\begin{array}{r}
$-1$
\end{array}$  &
$1/3$
  & 1/3 & $\begin{array}{r}
$1/2$
\end{array}$ & $\begin{array}{r}
$1/2$
\end{array}$ & $\begin{array}{r}
1/2
\end{array}$& $\begin{array}{r}
1/2
\end{array}$\tabularnewline
$t_{L\downarrow}^\dagger $ & $\begin{array}{r}
2/3
\end{array}$ & $\begin{array}{r}
$1$
\end{array}$  &
$1/3$
  & 1/3 & $\begin{array}{r}
$1/2$
\end{array}$ & $\begin{array}{r}
$1/2$
\end{array}$ & $\begin{array}{r}
$-1/2$
\end{array}$& $\begin{array}{r}
1/2
\end{array}$\tabularnewline

 $b_{R\uparrow}^\dagger $  &  $\begin{array}{r}
$-1/3$
\end{array}$ & $\begin{array}{r}
1
\end{array}$  &
$1/3$
  & -2/3 & $\begin{array}{r}
0
\end{array}$ & $\begin{array}{r}
0
\end{array}$ & $\begin{array}{r}
1/2
\end{array}$& $\begin{array}{r}
1/2
\end{array}$\tabularnewline
$b_{R\downarrow}^\dagger $  &  $\begin{array}{r}
$-1/3$
\end{array}$ & $\begin{array}{r}
$-1$
\end{array}$  &
$1/3$
  & -2/3 & $\begin{array}{r}
0
\end{array}$ & $\begin{array}{r}
$0$
\end{array}$ & $\begin{array}{r}
$-1/2$
\end{array}$& $\begin{array}{r}
1/2
\end{array}$\tabularnewline
$b_{L\uparrow}^\dagger $  &  $\begin{array}{r}
$-1/3$
\end{array}$ & $\begin{array}{r}
$-1$
\end{array}$  &
$1/3$
  & 1/3 & $\begin{array}{r}
$-1/2$
\end{array}$ & $\begin{array}{r}
$1/2$
\end{array}$ & $\begin{array}{r}
1/2
\end{array}$& $\begin{array}{r}
1/2
\end{array}$\tabularnewline

$b_{L\downarrow}^\dagger $  &  $\begin{array}{r}
$-1/3$
\end{array}$ & $\begin{array}{r}
1
\end{array}$  &
$1/3$
  & 1/3 & $\begin{array}{r}
$-1/2$
\end{array}$ & $\begin{array}{r}
$1/2$
\end{array}$ & $\begin{array}{r}
$-1/2$
\end{array}$& $\begin{array}{r}
1/2
\end{array}$\tabularnewline

$W^3_z$   &  $\begin{array}{r}
$0$
\end{array}$ & $\begin{array}{r}
$-1$
\end{array}$  &
$0$
  & $\begin{array}{r}
0
\end{array}$ & $\begin{array}{r}
0
\end{array}$ & $\begin{array}{r}
1
\end{array}$ & $\begin{array}{r}
$1$
\end{array}$& $\begin{array}{r}
1
\end{array}$\tabularnewline

$H_t $, \ \ \  $H_b^\dagger  $  &  $\begin{array}{r}
$0$
\end{array}$ & $\begin{array}{r}
--
\end{array}$  &
$0$
  & 1& $\begin{array}{r}
$-1/2$
\end{array}$ & $\begin{array}{r}
$1/2$
\end{array}$ & $\begin{array}{r}
$0$
\end{array}$& $\begin{array}{r}
0
\end{array}$\tabularnewline
\hline
\end{tabular}
\par\end{raggedright}
\caption{  Eigenvalues for t- and b-quark operators, and their combinations for   the neutral-vector   $W^3_z$, and   neutral-scalar $H$  components, defined respectively in Eqs. \ref{NeutralB} and  \ref{HiggsNorma}. The operators are: the electric charge $Q=I^3+\frac{1}{2}Y$, the momentum direction for the massless fermion  basis $\hat k$, the baryon number $B$, hypercharge $Y$, weak isospin component  $I^3$, weak isospin square    $\bar I^2$ ($I_s$ within $I_s(I_s+1)$), spin component along  $\hat z$ $S_z$, and total spin $\bar S^2$ ($S_s$ within $S_s(S_s+1)$). }
\end{table}
%
On Table 1 we describe these fermions written in terms of right-handed  ($R$) and left-handed components ($L$), with their quantum numbers whose operators are  presented  next.
%
 The relevant SM bosons and conserved quantities are bilinear components that can be written in this basis; we assume they are formulated after the Higgs mechanism.
  One-body operators  are constructed from matrix elements in a generic basis $|i\rangle$, and associated operators $a_i$,  as $Op\rightarrow \sum_{ij}\langle i |Op| j \rangle a_i^\dagger a_j $.  We write  the  operators that define these states and also  the SM vertices; some of  these operators  constitute symmetry generators obtained from
     conserved charges,  and we  list relevant ones:

   For the scalar  isospin   and hypercharge, the same separation can be made, where the sequence in  Eq. \ref{SVmass} contains the adjoint-representation fields $W_\mu^i$, and the  field $B_\mu$. Starting with the latter's associated hypercharge operator,
we  include a  flavor space for the t, b  quark pair:
 \begin{eqnarray}
 Y_o= \frac{4}{3} P_{t_R}-\frac{2}{3}P_{b_R}+ \frac{1}{3}(P_{t_L}+P_{b_L}),
 \end{eqnarray}
where, e.g,  $P_{t_L}=\sum_i    |t_L i \rangle \langle t_L i|$ a neutral component that sums over the spin elements generated by the   $\gamma_\mu$ in Eq. \ref{SVmass}. In its  second quantized form,
  \begin{eqnarray}\label {Yp}Y= \sum_{i} [ \frac{4}{3}t_{Ri}^\dagger t_{Ri}-\frac{2}{3}b_{Ri}^\dagger b_{Ri}+\frac{1}{3}(t_{Li}^\dagger t_{Li} +b_{Li}^\dagger b_{Li})]. \end{eqnarray}
 Other operators are given in such a form.  The SU(2)$_L$ generators $I^i$:
   \begin{eqnarray}\label {Is}I^3= \frac{1}{2}\sum_{i}  (t_{Li}^\dagger t_{Li}-  b_{Li}^\dagger    b_{Li}) , \end{eqnarray}
   \begin{eqnarray}\label {Ip}I^+=I^1+i I^2=   \sum_{i}      {b_{Li}} t_{Li}^\dagger     , \\
  I^-=I^1-i I^2=   \sum_{i}   t_{Li}  b_{Li}^\dagger  ,
 \end{eqnarray}
  satisfying $ [I^3,I^+]=I^+, $  $[I^3,I^-]=-I^-, $  $[I^+,I^-]=2I^3$,   $[Y,I^{\pm,3}]=0;$ the   baryon number
   \begin{eqnarray}\label {Ba}B=   \frac{1}{3}\sum_{i} (t_{Ri}^\dagger t_{Ri}+{b_{Ri}}^\dagger b_{Ri}+ t_{Li}^\dagger t_{Li} +b_{Li}^\dagger b_{Li}),   \end{eqnarray}  and     spin component along  $\hat z$
   \begin{eqnarray}\label {SpinZ}S_z=  \frac{1}{2} \sum_{qQ} ( q_{Q\uparrow}^\dagger q_{Q\uparrow}-q_{Q\downarrow}^\dagger q_{Q\downarrow}), \end{eqnarray}
   satisfying $[S_z,B]=0,$  $[S_z,Y]=0,$   $[B,Y]=0$, $[S_z,I^{\pm,3}]=0,$ $[Y,I^{\pm,3}]=0,$  $[B,I^{\pm,3}]=0$.

   Composite states may   also be  constructed that generalize Cooper pairs in superconductivity. In particular, bispinor operators describe bosons,  except for combinations of the form $q_{iL}^\dagger  q_{iR}^\dagger$  or  $q_{iL} q_{iR}, $ which cannot form a state, as   each term in the pair requires the same creation operator,   so that the non-vanishing operator acting  on   $ |0  \rangle$ is  squared, based on the massive-state  normalization of
 $q_{i}^\dagger|0  \rangle$,  $\bar q_{i}^\dagger|0  \rangle $.


 Similarly, the neutral  vector
 \begin{eqnarray}\label{NeutralB}W^3_z=     t_{L\uparrow}^\dagger t_{L\uparrow}
 -t_{L\downarrow}^\dagger t_{L\downarrow}
 -b_{L\uparrow}^\dagger b_{L\uparrow}+  b_{L \downarrow}^\dagger    b_{L \downarrow}\end {eqnarray}
  has quantum numbers as in Table 1. $W^3_z$ is derived from terms of the form  $\tau^3 W_\mu^3(x)$ in Eq.  \ref{SVmass},    shares   with   $ I^3$  weak isospin quantum numbers, and has also spin 1 under the Lorentz group.
  Regarding the freedom
   in defining boson operators as $I^i$  and $W_z^3$, we note we choose them so as to  reproduce  the  fermion coupling, and in particular, the  chirality property. Other components can be constructed, e. g., by applying step operators; these are presented in the Appendix.

The Higgs field is constructed so that gauge  invariance is satisfied, with quantum numbers common for the t and b field components:
 \begin{eqnarray}\label{HiggEpan} H_{ot} =\sum_i    |t_R i \rangle \langle t_L i|   \end{eqnarray}
with isospin-hypercharge $i^3=-1/2$, $y=1$, also manifest in   Eq. \ref{Htilde}.
A similarity transformation can be used to obtain the component:
\begin{eqnarray}\label{TildeHiggEpan}  H_{ob} =\sum_i    |b_R i \rangle \langle  b_L i|,   \end{eqnarray}  with   $i^3 =1/2$, $y=-1$.

The SM assumes conventionally a classical  procedure  for the vev;
it assumes action on fields is through a multiplicative constant.  Here we extend such an approach in that   relevant degrees of freedom are considered in the expectation value, departing from the same field arrangement.
   The main purpose of this calculation is to write SM vertices in this basis, and to  relate   the   vector-boson  masses in the scalar-vector (SV) vertex, and  fermion masses in the scalar-fermion (SF) Yukawa term.  Vector operators are constructed so as to match vertices, fermion chirality and coupling.

 Thus, the  electroweak   vertex operator in Eq. \ref {SVmass} that defines the mass-component Hamiltonian squared  contains the  scalar
 \begin{eqnarray}\label {HiggsNorma}H&=& \frac{v}{2} [\chi_t   (t_{R\uparrow}^\dagger  t_{L\uparrow} +t_{R\downarrow}^\dagger t_{L \downarrow})+\chi _b  (b_{L\uparrow}^\dagger  b_{R\uparrow} +b_{L\downarrow}^\dagger  b_{R\downarrow})] \\
  &=&\frac{v}{2} (  \chi_t H_t+\chi_b H_b^\dagger)
 ,  \end{eqnarray}
where  $\chi_t,$ $\chi_b $ are   parameters, and
   \begin{eqnarray} \label{HtOpeSQ} H_t &=& t_{R\uparrow}^\dagger  t_{L\uparrow} +t_{R\downarrow}^\dagger t_{L\downarrow} \\
 {H_b} &=& b_{R\uparrow}^\dagger  b_{L\uparrow} +b_{R\downarrow}^\dagger b_{L\downarrow} ,
  \label{HbOpeSQ}
\end{eqnarray}
    are associated to $ H_{ot}$,  $ H_{ob}$ in Eqs.  \ref{HiggEpan}, \ref{TildeHiggEpan}, respectively.
The freedom choice of $\chi_t,$ $\chi_b $  reflects an  extended parameter space in the scalar-vector term implying a symmetry is present, akin to custodial symmetry\cite{BesproRom}.

    \section{Particle Masses}

 Using the second-quantized fermion basis, we calculate the SM-vector masses.   From  Eqs. \ref{massquark}, \ref{massquarkdestru}, we find the conditions
   \begin{eqnarray}\label {HiggsSqExp}   \langle 0  |;H_q H _{q}^\dagger;| 0 \rangle  =
   2     \end{eqnarray}  (see Appendix), where we define the antinormal ordering as the operation that anticommutes creation operators to the right, subtracting the vev component:
      \begin{eqnarray}\label{AntiNormalOrder} ;a^\dagger   b;=   - b a^\dagger \  \  \  ;a   b^\dagger ;=    a b^\dagger\end{eqnarray}
   matching the classical normalization in Ref. \cite{BesproRom}.


\subsection{  W$^\pm$ mass}  We use the  spin-1  quantized component (see appendix)\footnote{The same calculation can be done for the opposite polarization $W_{-1}^+=\frac{1}{2}[W_x^1-i W_y^1+i (W_x^2-i W_y^2) ]$.}:  $W_1^+=\frac{1}{2}[W_x^1+i W_y^1+i (W_x^2+i W_y^2)]=2 t^\dagger_{L\uparrow}b_{L\downarrow}$,
which  reproduces $\tau^+\sigma^+$, $\tau^+=\tau^1+i\tau^2$, $\sigma^+=\sigma^1+i\sigma^2$,  and $\boldmathsigma$ represent the spin-associated Pauli matrices.
 $[ H,W_1^+] = 2(\chi_t H_t^1 -\chi_b {H_b^1}^\dagger) $, where
 $ H_t^1=   t^\dagger_{R\uparrow}b_{L\downarrow},$
 $  {H_b^1}^\dagger=t^\dagger_{L\uparrow}b_{R\downarrow}  $
leading to (see Appendix)
\begin{eqnarray}\label{Wmass}\langle 0|;[ v H, \frac{1}{2} g W_1^+]^\dagger[ v H , \frac{1}{2} g W_1^+];| 0 \rangle=(|{\chi_t}|^2+|{\chi_b}|^2+\chi_t^*\chi_b+\chi_t\chi_b^*)\frac{1}{4}g^2 v^2.\end{eqnarray}

Coincidence with the classical result for the W mass in Eq. \ref{particleMasses} requires
 \begin{eqnarray}\label {NormaConditionGeneral}
 |{\chi_t}|^2+|{\chi_b}|^2+\chi_t^*\chi_b+\chi_t\chi_b^*=1.\end{eqnarray}
 Assuming $\chi_t$ real, $\chi_b$ and imaginary, the cross term is eliminated, implying the
restriction in coincidence with the result with a spin basis\cite{BesproRom}.
 \begin{eqnarray}\label {NormaCondition}|{\chi_t}|^2+|{\chi_b}|^2 =1.\end{eqnarray}.


\subsection{   Z mass} The neutral  second-quantization operator in Eqs. \ref{Yp}, \ref{Is}   have $H$ as eigenoperator:
$[      {g }^2    I^3-\frac{1}{2}{g'}^2 Y ,  H]= - \frac{1}{2}(g^2+g'^2) H$ so
 \begin{eqnarray}\label {NeutralMassOperator}
  \frac{1}{({g }^2  +{g'}^2)}\langle 0 |;[    {g }^2  I^3-\frac{1}{2}  {g'}^2 Y , v H]^\dagger[  {g }^2   I^3-\frac{1}{2}{g'}^2 Y ,v H];| 0 \rangle=\\
  (|{\chi_t}|^2+|{\chi_b}|^2+\chi_t^*\chi_b+\chi_t\chi_b^*)(g^2+g'^2) v^2 /8 , \nonumber
  \end{eqnarray}
with SM consistency requiring condition in Eq. \ref{NormaConditionGeneral}.

Eq. \ref{NeutralMassOperator} is consistent with the expectation value of a normalized configuration composed of a combination of fields with
   the  coupling constant interpreted as   normalization\cite{besproCoupling,bespro9m1};
thus, $\frac{1}{2}gI^3|0  \rangle$ is the state associated to W$^3_0$ with normalization $ \langle  0| \frac{1}{4} I^3  I^3|0  \rangle = N,  $ $g=2/\sqrt{N} $,  where   orthogonality   eliminates the cross terms.

     In comparison,   $H_{oq}$ in Eqs. \ref{HiggEpan}, \ref{TildeHiggEpan}, has  normalized expressions\cite{BesproRom} $\frac{1}{\sqrt{2}}H_{oq}$, $\frac{1}{  2 }(H_{oq}+H_{oq}^\dagger)$,  while $H_{q}$, in Eqs. \ref {HtOpeSQ}, \ref{HbOpeSQ} are normalized as $\sqrt{2}H_{q}$, $ H_{q}+H_{q}^\dagger $.

 Thus, the Higgs component defines also the   mass operator
   \begin{eqnarray}\label {HiggsYukawa}H_m=\frac{v}{\sqrt{2}}[\chi_t (t_{R\uparrow}^\dagger  t_{L\uparrow} +t_{R\downarrow}^\dagger t_{L\downarrow})+\chi_b  (b_{L\uparrow}^\dagger  b_{R\uparrow} +b_{L\downarrow}^\dagger  b_{R\downarrow})],  \end{eqnarray}
 with the normalization set by the property\footnote{The phase can be fixed by redefining $H _{q}$.} \begin{eqnarray}
\label{VacuumExpHq} \langle 0  | \frac{v}{\sqrt{2}} H _{q} | 0 \rangle  =
  -\frac{v}{\sqrt{2}}.  \end{eqnarray} $H_m$ reproduces the mass relation:
 $  \langle 0|;[ H_m, \frac{g}{ 2 }(W_1^++  W_1^-)]^\dagger[ H_m,\frac{g}{ 2 }(W_1^++  W_1^-)];| 0 \rangle=(|{\chi_t}|^2+|{\chi_b}|^2+\chi_t^*\chi_b+\chi_t\chi_b^*)\frac{1}{4}g^2 v^2$
     (see Appendix).

\subsection{  Top, bottom quark masses}  In turn, $H_m$ is the resulting Yukawa operator that gives mass to fermions    \begin{eqnarray}
\label{HiggsYukawaNOrder}:H_m+ H_m^\dagger:=\frac{v}{\sqrt{2}}\sum_{i} [ \chi_t (t_{i}^\dagger t_{i}+\bar t_{i}^\dagger \bar t_{i})+  \chi_b ( b_{i}^\dagger   b_{i}+\bar b_{i}^\dagger \bar   b_{i})]. \end{eqnarray}
$H_m$ reproduces the  mass relations
\begin{eqnarray}\label {Hm}[H_m+ H_m^\dagger,t_i^\dagger]=\frac{v \chi_t}{\sqrt{2}} t_i^\dagger=m_t t_i^\dagger,
\end{eqnarray}
\begin{eqnarray}\label{FermionMassGiving}
  [H_m+ H_m^\dagger,b_i^\dagger]=\frac{v \chi_b}{\sqrt{2}} b_i^\dagger=m_b b_i^\dagger,
   \end{eqnarray} (and corresponding rules for annihilation operators,) where  $\chi_t$,  $\chi_b$   are interpreted as  the t, b Yukawa coefficients.



  We shall use  the demand of coincidence with standard model to limit $\chi_{t,b}$.
Eqs. \ref{Hm}, \ref{FermionMassGiving} are compared to the   mass-giving for a vector  in Eq. \ref{NormaConditionGeneral}.
   Given the assumption of the same underlying operator, we associate   generic scalars  in Eqs.  \ref{HiggsNorma},  \ref{HiggsYukawa},
      implying for the fermion masses
  \begin{eqnarray}\label {MassCondition}{m_t}^2+{m_b}^2=v^2/2,\end{eqnarray}
which reproduces the relation in Ref.  \cite{BesproRom}, where the quantum extension encompasses a phase parameter from   Eq. \ref{NormaConditionGeneral}.

\subsection{Composite-scalar vacuum expectation value}
 The Higgs component  $H_m$ in Eq.  \ref {HiggsYukawa} obtains the vev, using
 Eq. \ref {VacuumExpHq},
\begin{eqnarray}
\label{VacuumExpChis} \langle 0  | H_m | 0 \rangle  = - \frac{v}{\sqrt{2}}(\chi_t +\chi_b),
\end{eqnarray} assuming for the Yukawa coefficients the assumed  polar form $|\chi_t| e^\theta $ and $|\chi_b|$, with   $\theta$ the relative phase.
The demand that $\langle 0  | H_m | 0 \rangle= - \frac{v}{\sqrt{2}}$ leads to   \begin{eqnarray}\label{ConstrainChitChib}  \chi_t +e^{i\theta} \chi_b  =1.\end{eqnarray}
 This implies the quantized contribution  \begin{eqnarray}
\label{Chi}  \chi =\sqrt{|\chi_b|^2+|\chi_t|^2+ |\chi_b \chi_t| (e^{i\theta}+e^{-i\theta})},  \end{eqnarray}  satisfies the normalization in Eq.   \ref{NormaConditionGeneral}; in other
words, the classical W, Z masses are consistent with a quantum mechanically constrained vev.


From the classical normalization condition\cite{BesproRom}, \begin{eqnarray}\label{ClassicalNormalization}|\chi_t| =\sqrt{1-  |\chi_b|^2}\end{eqnarray} is   substituted into $\chi$ in Eq. \ref {Chi}, and the latter is plotted  in Fig. 1.  Interestingly, the correct $\chi$ near 1 values constrain   $|\chi_b|$    to small  values, more likely in   the  denser $\theta$ region,
 as it approaches $ \pi/2$  (a similar condition is obtained for  $\pi/2\leq\theta\leq \pi$), predicting the $m_t$, $m_b$ mass hierarchy;
thus,  $|\chi_b|\ll 1$, (or, less likely, $|\chi_b|\sim 1$), so  Eq. \ref{NormaConditionGeneral} reproduces  the $m_t\sim v/\sqrt{2} $ mass prediction.
%
\begin{figure}
\begin{centering}
\includegraphics[scale=0.7]{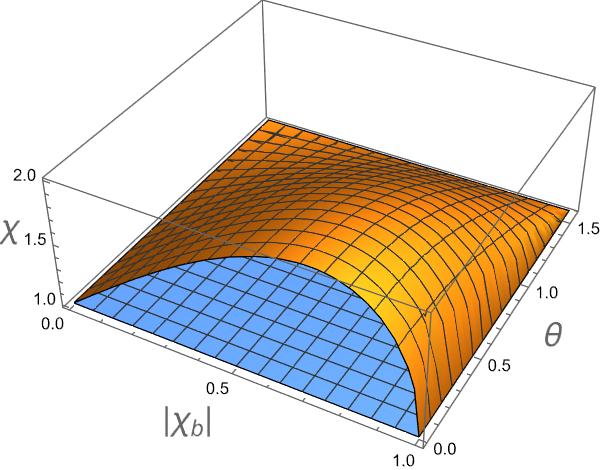}
\par\end{centering}
\caption{ $  \chi$ component in Eq. \ref{Chi},  compared with  expected value 1,     as obtained in $M_Z$, $M_W $,  Eqs. \ref{Wmass},        \ref{NeutralMassOperator}, constrained by Eq. \ref {ConstrainChitChib}, as function of $\theta$ phase and $\chi_t$, favoring, e. g.,  $0\leq|\chi_b|\ll 1$, and so $0 \ll   |\chi_t|\leq 1$.  }

%
 \label{FeynmannDiag}
\end{figure}

\section{Conclusions}

 This paper deals with the low-energy regime in which   the scalar Higgs field acquires a vev, generating mass upon itself and the other fields. Boson fields are written  in a composite
description,  with  a chiral fundamental-representation fermion  basis producing vectors in  the adjoint one, and  reproducing the SM\footnote{By construction, the Higgs particle remains in the weak doublet representation.}. Their associated SM quantum numbers  support  such  connections.

  Effectively, we  focus  on the scalar-vector and scalar-fermion vertex  mass components, providing  the vector and fermion masses. Within such vertices, the field's spatial component is factored out, so we concentrate on discrete   spin-isospin degrees of freedom, relevant in mass generation. A common mass-giving scalar element appears in each vertex  that identifies    the scalar field in the two expressions. The second-quantized   SM W and Z masses
   in Eqs. \ref{Wmass},  \ref{NeutralMassOperator},  respectively,
  are  reproduced under the normalization condition   \ref{NormaConditionGeneral}, which restricts quark masses, with    Eq. \ref{MassCondition} as particular case,  from shared quantum numbers of scalar generators with  $\chi_b$, $\chi_t$ parameters; under a hierarchy condition, $m_t\sim v/\sqrt{2} $ is reproduced. The vev normalization condition in Eq. \ref{VacuumExpHq}  implies  Eq. \ref{NormaConditionGeneral};   demanding the classical normalization in Eq. \ref {ClassicalNormalization}, one derives general hierarchy condition  for $m_t$ and $m_b$, as shown in Fig. 1. Based on current values of $m_t=172-176$ GeV, extracted kinetically or through pole methods\cite{PDG}, this relation is satisfied with a .6\% accuracy.



 In comparison, Ref. \cite{BesproRom} constructs an equivalent spin  basis, departing  from a SM extension,  within a classical description; it also digresses on applying the same scheme to    the lower-mass quarks, although    they are irrelevant  at  $v$ scales. Consistency with the W, Z masses imposes a normalization for Ref. \cite{BesproRom} for the arbitrary Yukawa parameters  $\chi_b$, $\chi_t$  with the freedom of choice  within  a symmetry.

  Ref. \cite{BesproRom} applies  a Clifford algebra to describe fermions and boson degrees of freedom within a
     SM Lagrangian operator  and state  representation;   here,  discrete degrees  of freedom are separated  from such an algebra, and the vev   emerges from  quantized   operators,  which  supports and complements  the   Ref. \cite{BesproRom} framework.

The fields' fermion   expression  encompasses two interpretations: a formal one, as a   basis  reflecting the SM's composite structure  or as a physical one, suggesting  common dynamics, akin as pair behavior in   superconductivity theories\cite{Schrieffer}. The operators can be  also interpreted as a model on its own.




The usual SM description  expands  classically   the  scalar around the vev
   at low  energy,     extracting the fields' mass.  We  consider  quantum aspects, concentrating on  the  spin, isospin  operators   involved,  with new information obtained: a  mass relation, with a  phase connection.

Thus,  this work's quantum approach can enrich   SM extensions that connect the t- b-quarks  with a scalar Bose-Einstein condensate.
Quantization applications   provide SM information, as this
and other work\cite{Wetterich} use  compositeness,  complementing other methods: through monopoles\cite{ellis}, gravity\cite{Radenovic},  anomaly cancellation and supergravity\cite{WenYin}, gauge invariance\cite{DiStefano},  and spin\cite{Mankoc,BesproIni,Besprosub,besproCoupling,bespro9m1}. Further links  and connections  in    phenomenological constants  within the SM will enhance these   beyond-the-SM frameworks.

In conclusion, independently of whether compositeness is  formal or physical, the mathematical second-quantized fermion-basis fields' presentation    expresses SM properties,  as heavy-particle degrees of freedom can be described in simple elements.

\section*{Appendix}


\renewcommand{\theequation}{A\arabic{equation}}
 \setcounter{equation}{0}

 {\bf $\gamma_5$ operator}

 This pseudoscalar operator is relevant, as the electroweak interactions are chiral.
It is written in second-quantized form, in terms of massive fermionic operators (using one mode):
\begin{eqnarray}\label{Gamm5}
\hat \gamma_5&=&\sum_q ( q_\uparrow^\dagger \bar q_\downarrow^\dagger  +\bar q_\downarrow   q_\uparrow + q_\downarrow^\dagger \bar q_\uparrow^\dagger  +\bar q_\uparrow  q_\downarrow)
,
\end {eqnarray}
with four eigenvalues as
 \begin{eqnarray}\label{massquarkApp}q_{R\uparrow}^\dagger= \frac{1}{\sqrt{2}}( q_{\uparrow}^\dagger + \bar q_{\downarrow} ) \end {eqnarray}
  \begin{eqnarray}\label{massquarkdestruApp}  q_{L\uparrow}^\dagger = \frac{1}{\sqrt{2}}(  q_{\uparrow}^\dagger - \bar  q_{ \downarrow }),\end {eqnarray}
obtained by inverting Eqs. \ref{massquark}, \ref{massquarkdestru}, and  satisfying canonical anticommutation relations.
In terms of these chiral operators,
\begin{eqnarray}\label{Gamm5RL}
\hat \gamma_5=\sum_i ( q_{R i}^\dagger q_{R i}  -q_{L i}^\dagger q_{L i}).
\end {eqnarray}

Vevs  are obtained for
  \begin{eqnarray}\label{DefineHdaggeri} H_i  =  q_{Ri}^\dagger  q_{Li},
      \end{eqnarray}  written in terms of such chiral fermion operators:
   \begin{eqnarray}\label{VacuumExpHpHdag}\langle   H_\uparrow+  H_\uparrow^\dagger  + H_\downarrow+  H_\downarrow^\dagger\rangle & = &
    \langle  q_{\uparrow}^\dagger    q_{\uparrow} -\bar  q_{\downarrow}   \bar q_{\downarrow}^\dagger +  q_{\downarrow} ^\dagger     q_{\downarrow}  - \bar q_{\uparrow}    \bar q_{\uparrow}^\dagger  \rangle \\
   & = &\langle  q_{\uparrow}^\dagger    q_{\uparrow} +  \bar q_{\downarrow}^\dagger \bar  q_{\downarrow}  +  q_{\downarrow} ^\dagger     q_{\downarrow}  +     \bar q_{\uparrow}^\dagger \bar q_{\uparrow} -2 \rangle,
     \end {eqnarray}
implying Eq. \ref{VacuumExpHq}.
To calculate  $\langle H ^\dagger  H  \rangle$, $H$ in Eq. \ref{HiggsNorma}, its components satisfy, e. g.,
\begin{eqnarray}\label{CalcHdaggerH}
\langle ; (  q_{L\uparrow}^\dagger q_{R\uparrow} )^\dagger        q_{L\uparrow}^\dagger  q_{R\uparrow} ; \rangle=\langle; q_{R\uparrow}^\dagger q_{L\uparrow}    q_{L\uparrow}^\dagger  q_{R\uparrow} ; \rangle=  \langle
\bar q_\downarrow    q_\uparrow   q_\uparrow ^ \dagger \bar q_\downarrow  ^\dagger \rangle=1,
 \end {eqnarray}
while for cross terms
\begin{eqnarray}\label{CalcHdaggerHCross}
\langle ;(  q_{R\uparrow}^\dagger q_{L\uparrow} )^\dagger        q_{L\downarrow}^\dagger  q_{R\downarrow};  \rangle=\langle ; q_{L\uparrow}^\dagger q_{R\uparrow}    q_{L\downarrow}^\dagger  q_{R\downarrow} ; \rangle=0,
\end {eqnarray} which lead to Eq. \ref{HiggsSqExp}.

We provide the second quantized $W_\mu^i$ (labeled by $\hat\tau^i$):

\begin{eqnarray}\label {IsW}W^3_0&=& \hat\tau^3=   \sum_{i}  (t_{Li}^\dagger t_{Li}-  b_{Li}^\dagger    b_{Li}) , \end{eqnarray}
   \begin{eqnarray}\label {WIp}W^+_0&=& \hat\tau^+=\frac{1}{\sqrt{2}}( \hat \tau^1+i \hat\tau^2)= \sqrt{2}  \sum_{i}      {b_{Li}} t_{Li}^\dagger     , \\
  W^-_0&=& \hat\tau^- =\frac{1}{\sqrt{2}}(\hat \tau^1-i \hat\tau^2)= \sqrt{2}   \sum_{i}   t_{Li}  b_{Li}^\dagger  ,
 \end{eqnarray}

\begin{eqnarray}\label {IsWx}W^3_x &=&    t_{L \uparrow}^\dagger t_{L\downarrow}+t_{L \downarrow}^\dagger t_{L\uparrow}- b_{L \uparrow}^\dagger    b_{L\downarrow}- b_{L \downarrow}^\dagger    b_{L\uparrow})   , \end{eqnarray}
    \begin{eqnarray}\label {Ipx} W^+_x&=& \sqrt{2} (  t_{L\uparrow}^\dagger {b_{L\downarrow}} +       t_{L\downarrow}^\dagger  {b_{L\uparrow}} )   \\
  W^-_x &=&  \sqrt{2} (   b_{L\uparrow}^\dagger t_{L\downarrow}+  b_{L\downarrow}^\dagger t_{L\uparrow}  ),
 \end{eqnarray}

\begin{eqnarray}\label {IsWy} W^3_y&=&   -i    ( t_{L\downarrow}t_{L \uparrow}^\dagger - t_{L\uparrow} t_{L \downarrow}^\dagger-     b_{L\downarrow} b_{L \uparrow}^\dagger+     b_{L\uparrow} b_{L \downarrow}^\dagger )  ,   \end{eqnarray}
   \begin{eqnarray}\label {Ipy}W^+_y&=&   -i\sqrt{2}(  t_{L\uparrow}^\dagger {b_{L\downarrow}} -     t_{L\downarrow}^\dagger  {b_{L\uparrow}} )       , \\
 W^-_y&=&    \  -i \sqrt{2}(  b_{L\uparrow}^\dagger t_{L\downarrow} -   b_{L\downarrow}^\dagger t_{L\uparrow} )    ,
 \end{eqnarray}

\begin{eqnarray}\label {IsWz} W^3_z &=&      t_{L \uparrow}^\dagger t_{L\uparrow}-t_{L \downarrow}^\dagger t_{L\downarrow} - b_{L\uparrow}^\dagger    b_{L\uparrow} +b_{L\downarrow}^\dagger    b_{L\downarrow} , \end{eqnarray}
   \begin{eqnarray}\label {WIpz}W^+_z &=&     \sqrt{2}(   t_{L\uparrow}^\dagger  {b_{L\uparrow}}- t_{L\downarrow}^\dagger  {b_{L\downarrow}}  )   , \\
 W^-_z&=&   \sqrt{2}(   b_{L\uparrow}^\dagger {t_{L\uparrow}}-  b_{L\downarrow}^\dagger {t_{L\downarrow}} )
 \end{eqnarray}
The $W$-mass calculation uses  $ W^+_1=\frac{1} {\sqrt{2}}( W^+_x+i W^+_y)=2 t^\dagger_{L\uparrow}b_{L\downarrow}$.

The expectation value with $H$ uses:
$[ H,W_1^+] =2( \chi_t H_t^1 -\chi_b {H_b^1}^\dagger) $, with  $ H_t^1=   t^\dagger_{R\uparrow}b_{L\downarrow},$
 $  {H_b^1}^\dagger=t^\dagger_{L\uparrow}b_{R\downarrow}  $. For direct terms,  e. g.,
\begin{eqnarray}\label{CalcHdaggerHW}
\langle ; (  t_{L\uparrow}^\dagger  b_{R\downarrow} )^\dagger        t_{L\uparrow}^\dagger  b_{R\downarrow} ;  \rangle=\langle ; b_{R\downarrow}^\dagger t_{L\uparrow}    q_{L\uparrow}^\dagger  q_{R\uparrow} ; \rangle=   
0,
 \end {eqnarray}
while for cross terms,
\begin{eqnarray}\label{CalcHdaggerHCrossW}
\langle ;( b_{R\downarrow}^\dagger t_{L\uparrow}  )^\dagger         t_{R\uparrow}^\dagger  b_{L\downarrow} ; \rangle=1,
\end {eqnarray} which lead to Eq. \ref{Wmass}.

\baselineskip 31pt\vfil\eject 

{\bf Acknowledgements}
 The
authors thank Dr.  Jose Wudka for discussions, and   support from  DGAPA-UNAM through projects IN112916, IN117020, IN112822.

\end{document}